# Assessing the Dosimetric Effects of High-Z Titanium Implants in Proton Therapy Using Pixel Detectors

**C. Bălan**[a,b], **C. Granja**[c], **G. Mytsin**[d], **S. Shvidky**[d], **A. Molokanov**[d], **V. Chiş**[a] and **C. Oancea**[c,*]

[a] *Faculty of Physics, Babeş-Bolyai University, Cluj-Napoca, Romania*
[b] *Department of Radiotherapy, The Oncology Institute 'Prof. Dr Ion Chiricuţă', Cluj-Napoca, Romania*
[c] *ADVACAM, Prague, Czech Republic*
[d] *International Intergovernmental Organization Joint Institute for Nuclear Research, Dubna*

*E-mail*: cristina.oancea@advacam.cz

ABSTRACT:

A rapid increase in the use of proton therapy for cancer treatment has been seen in the last decade due to its clinical advantages. Therefore, more and more patients with implants and other metallic devices will be among those who will be treated. This study experimentally examines the effect and changes in the delivered fields, using water-equivalent phantoms with and without titanium (Ti) dental implants positioned along the primary beam path. We measure in detail the composition and spectral-tracking characterization of particles generated in the entrance region of the Bragg curve using high-spatial resolution, spectral and time-sensitive imaging detectors with a pixelated array provided by the ASIC chip Timepix3. A 170 MeV proton beam was collimated and modulated in a polymethyl methacrylate (PMMA). Placing two dental implants at the end of the protons range in the phantom, the radiation was measured using two pixeled detectors with Si sensors. The Timepix3 (TPX3) detectors equipped with silicon sensors measure in detail particle fluxes, dose rates (DR) and linear energy transfer (LET) spectra for resolved particle types. Artificial intelligence (AI) based-trained neural networks (NN) calibrated in well-defined radiation fields were used to analyze and identify particles based on morphology and characteristic spectral-tracking response. The beam was characterized and single-particle tracks were registered and decomposed into particle-type groups. The resulting particle fluxes in both setups are resolved into three main classes of particles: i) protons, ii) electrons and photons iii) ions. Protons are the main particle component responsible for dose deposition. High-energy transfer particles (HETP), namely ions exhibited differences in both dosimetric aspects that were investigated: DR and particle fluxes, when the Ti implants were placed in the setup. The detailed multi-parametric information of the secondary radiation field provides a comprehensive understanding of the impact of Ti materials in proton therapy.

# Contents

## 1. Introduction

Modern treatments in particle therapy are focused on adapting the particle beams to deliver optimal dose directly to the tumour, with a significant reduction of dose in healthy tissue [1–3]. Clinical advantages of proton radiotherapy are strongly associated with reduced absorbed dose in organs that are surrounding the irradiated volume [4–7]. The correlation between lower doses of radiation and the radiobiological effectiveness of protons in cancer cells promotes particle therapy as a novel approach in cancer treatment [8, 9]. Proton therapy facilities are growing fast, and more patients are referred to proton radiotherapy [10, 11]. In certain cases where metallic inserts (e.g., dental or spinal implants, hip replacements) made from high-Z materials are present, and particle therapy is medically indicated, it is necessary to consider all possible consequences, including imaging artifacts and dose perturbations caused by these inhomogeneities, before starting treatment [12–14].

Few studies have investigated the impact of metallic implants in particle therapy when located close to the tumor [12–17]. Titanium implants can lead to an increase in secondary particles during proton therapy because of nuclear interactions. When the high-energy protons interact with the titanium atoms in the implant, nuclear reactions can occur, resulting in the production of secondary particles, including heavier ions. These secondary ions can contribute to the overall dose rate and have different ranges and energy deposition patterns compared to the primary protons. This phenomenon can significantly alter the expected dose distribution, especially in tissues surrounding the implant. While most of the studies are evaluating the scattered dose resulted from protons interaction with artificial implants close to the metal-phantom interface or dose perturbation produced in the target volume, some of them reported changes in the linear energy transfer (LET) values of scattered particles [15–17]. This study is filling the gaps by providing a dosimetric characterisation of primary radiation from the Bragg curve, transversing Titanium (Ti) implants. High-resolution detectors being able to evaluate particles with a wide range of LET values starting 0.1 keV/µm were used. Taking advantage of the new generation of detectors, primary and secondary radiation produced in this experiment could be visualized by performing decomposition of the mixed particles field [18, 19]. With particle classification into different groups (protons, electrons and photons, ions with fast neutrons) dosimetric analysis through dose rates, particle fluxes and LET measurements was performed to quantify the impact of Ti implants in proton therapy.

This study aims to provide a comprehensive characterization of radiation produced by a clinical proton beam when two Ti implants are positioned directly into the beam's path, in the entrance region of the Bragg curve. Using high-resolution silicon (Si) sensors of the Timepix3 (TPX3) [20] detectors mounted near polymethyl methacrylate (PMMA) phantom particle flux, dose rate and LET with single-particle identification were conducted to quantify the influence of Ti dental implants.

## 2. Materials and Methods

### 2.1 Experimental setup using proton beam, PMMA phantom, and Ti implants

The experimental measurements were performed using a therapeutic proton beam from the synchrocyclotron at the Medico-Technical Complex of the Joint Institute for Nuclear Research (JINR). With a mean energy of 172.5 ± 2.7 MeV, up to $2 \cdot 10^6$ particles·$cm^2 \cdot s^{-1}$ were collimated in one 2 x 2 $cm^2$ field-size proton beam [21]. Two Ti dental implants, similar to those used in molar replacement, were attached at the edge of one 140 mm water equivalent PMMA block and placed in the beam's path in the entrance of the Bragg curve region [22, 23]. Figure 1 shows the experimental setup used in this study.

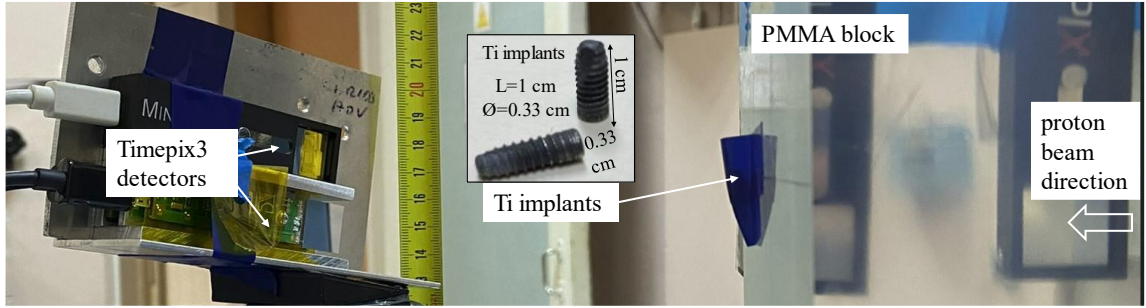

**Figure 1**. Irradiation geometry: A PMMA phantom was placed along the collimated 170 MeV proton beam (incident from right). Two Ti implants were attached to the edge of 140 mm $H_2O$ equivalent phantom. The pixel detectors were placed 14 cm behind the phantom along the beam axis to image radiation produced including the secondary component from the dental implants.

### 2.2 Timepix3 detectors. Readout methods, particle discrimination algorithms

The mixed field of radiation produced behind the phantom with Ti implants was measured with two Minipix Timepix3 detectors equipped with silicon sensors of 300 and 500 µm mounted at the edge of the PMMA phantom [24]. Figure 1 shows the Timepix3 detectors used which have a pixel matrix of 256 × 256 of total active area 14 × 14 $mm^2$. Data presented in this paper consists of events from the direct interaction of particles with the Si sensor collected in the region of 4.5 x 14 $mm^2$ from the region with implants and region of phantom without implants. Both energy and time of arrival were simultaneously registered by the pixeled array and sent to the ASIC chip. Data were acquired using the Pixet Pro software package v.1.8.0 (Advacam). Raw data were processed offline with the data processing engine (DPE) [18] called "*TraX Engine*". The spectral-tracking morphology of the single particle tracks is analyzed with pattern recognition algorithms and extensive calibrations performed in well-defined radiation fields [25]. Detailed particle-type classification includes artificial intelligence (AI) single-layer neural networks (NN) [18, 23].

## 3. Results and Discussion

### 3.1 Particle-type classification, event discrimination

The composition of primary and secondary radiation created along the Bragg curve can be decomposed into three groups of particle types: protons, electrons together with photons (X-rays and low-energy gamma rays), and ions (HETP tracks besides proton-induced events). Figure 2 shows 2D decomposition with each setup: without the metallic implant (left) and with the Ti implant (right). Figure 2a shows the 2D integrated energy deposition for 3000 events detected in an 8-second interval for each setup. The radiation field is further decomposed into three particle groups. **Protons** (Fig. 2b): the proton component of the radiation shows minimal variation between the two setups, with 1673 proton events detected without the implant and 1607 with the implant. **Electrons and Photons** (Fig. 2c): similarly, the electron and photon group exhibits minor differences between the setups, with 1324 events recorded without the implant and 1356 with the implant. **Ions** (Fig. 2d): a significant difference is observed in the ion component. With the titanium implant, the presence of ions increases, with 37 ions detected compared to only 3 in the setup without the implant.

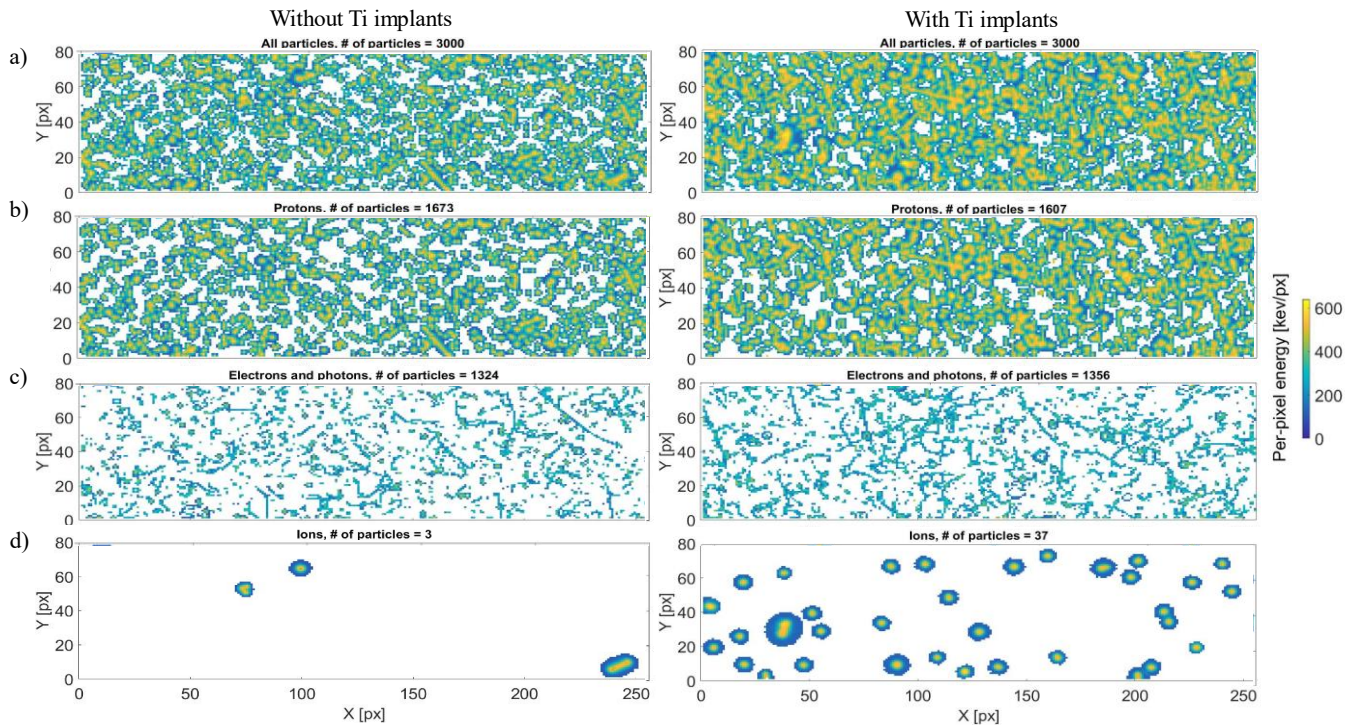

**Figure 2**. Decomposition of the mixed radiation field behind the phantom measured with the Timepix3 detector with Si sensors (300 and 500 µm thickness) for two cases: without Ti implants (left column) and with Ti implants placed in the primary proton beam. a) 2D imaging of integrated deposited energy produced by 3000 events measured in 8 s in both setups. Decomposition into three groups: b) protons; c) electrons & photons; and d) ions.

## 3.2 Particle fluxes and dose rates of scattered radiation

The proposed dosimetric analysis consists in a high-resolution investigation of different types of particles measured in two scenarios: with Ti dental implants and without them in a collimated proton incident beam in terms of particle fluxes, dose rates (DR) and LET spectra.

Figure 3 presents the mean values of particle fluxes (a) and mean DR (b) for all particles in black, protons in red, electrons together with photons in blue and ions in green. Both dosimetry parameters were mediated over 200 s of continuous proton beam with metallic inserts and without them respectively. Analyzing the fluxes, up to $550 \pm 40$ particles·cm$^{-2}$·s$^{-1}$ (with Ti) and 593 particles·cm$^{-2}$·s$^{-1}$ (without Ti) were detected in both cases by both TPX3 detectors used in this experiment. More protons were detected in the scattered radiation when the metallic inserts were removed from the beam's path, $337.3 \pm 22$ vs $294.9 \pm 22$ particles·cm$^{-2}$·s$^{-1}$ as a result of nonelastic multiple nuclear interactions of incident protons with Ti atoms [27]. For electrons and photons, there is no significant shift in their fluxes when Ti implants are placed in the beam, approx. 250 particles·cm$^{-2}$·s$^{-1}$. The situation is completely different for ion's contribution to the secondary generated radiation. Only $0.2 \pm 0.6$ particles·cm$^{-2}$·s$^{-1}$ were identified if no metals were present in the field, but their impact becomes statistically considerable in the other setup when almost 6 particles·cm$^{-2}$·s$^{-1}$ were reported. Overall, the flux of ions is very small.

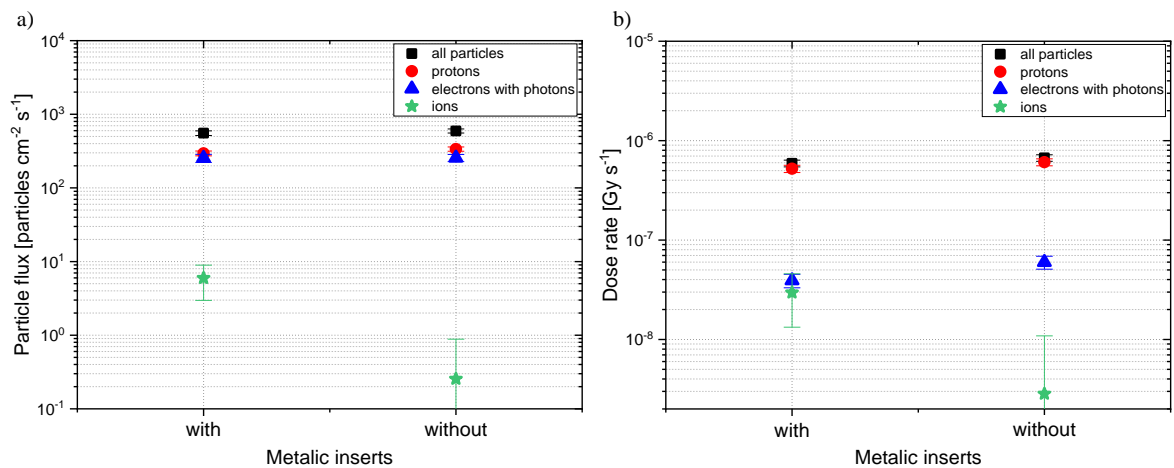

**Figure 3**. Partial particle fluxes (a) and dose rates (b) for the particle-type groups resolved in the scattered radiation in both setups: with and without Ti implant. Results were measured in 200 s. Statistical deviations are represented by the first level of standard deviation (k=1).

Figure 3b shows the dose rate of each group of particles. Protons remain the main group of particles responsible for dose deposition along the Bragg curve, regardless of the presence of the metallic component in the beam's path with a DR of $5.22 \cdot 10^{-7}$ Gy·s$^{-1}$ (with Ti) and $6.07 \cdot 10^{-7}$ Gy·s$^{-1}$ (without Ti). The presence of titanium implants led to a reduction in the contribution of electrons and photons to the final dose rate, lowering it to $0.39 \cdot 10^{-7}$ Gy·s$^{-1}$. This corresponds to a reduction of approximately 34% compared to the dose rate measured in the setup without titanium implants, which was $0.59 \cdot 10^{-7}$ Gy·s$^{-1}$.

The impact of the metallic inserts is highlighted by the ion contribution to the secondary radiation's DR, $0.29 \cdot 10^{-7}$ Gy·s$^{-1}$, being ten times higher than the DR measured in the same setup, but without metals which was $0.02 \cdot 10^{-7}$ Gy·s$^{-1}$. Although there is a substantial increase in the ion DR of the secondary radiation field, the DR is still very small compared to the dose of primary protons. The presence of titanium implants in the proton beam path modifies the composition of ions in DR. While the ion dose rate increases ten times in the presence of titanium, rising from $0.02 \cdot 10^{-7}$ Gy·s$^{-1}$ without the implant to $0.29 \cdot 10^{-7}$ Gy·s$^{-1}$ with the implant, this contribution remains relatively minor in the context of the overall dose distribution. The ion dose rate with the titanium implant constitutes only about 5.6% of the dose rate contributed by protons. In the absence of the implant, the ion dose rate represents an even smaller fraction, merely 0.3% of the proton dose rate. These results suggest that despite the noticeable increase in the ion component due to the presence of titanium, the primary protons remain the dominant contributors to the total dose deposition. This shows the complexity of dose distribution in proton therapy, particularly in the presence of high-Z materials where there is a need for careful consideration of secondary ion production in treatment planning.

### 3.3 LET spectra

The LET quantification is a key quantity in treatment planning used to minimize damage to healthy tissue surrounding the target volume [28]. In this study, detailed calculations were performed to analyze the directional and spectral aspects responsible for the LET distribution of primary and secondary radiation, as shown in Figure 4. The LET spectra for the mixed radiation field produced by a 170 MeV proton beam along the Bragg curve were measured using Timepix3 detectors with silicon sensors (LET$_{Si}$). Timepix-based detectors are suitable for LET detection on a large spectrum of energies and particle types [19, 29-32]. The particle's path into the Si sensor was correlated with the thickness of each detector that was used in measurements [26]. LET spectra is given in Figure 4 in dark blue for setup with Ti implants and light blue for the setup without them. Having the decomposition of the radiation field, the LET measured in Si (LET$_{Si}$) is presented separately for each group of particles.

The LET of protons exhibits a broad spectrum, ranging from 0.5 up to 8.5 keV·µm$^{-1}$. In the absence of Ti in the proton beam path, scattered protons exhibit a distinct LET peak around 1.5 keV·µm$^{-1}$. Multiple Coulomb scattering together with nonelastic nuclear interactions as the predominant interaction of protons with matter, could be responsible for the specific peak with a maximum value at 0.75 keV·µm$^{-1}$ [27]. Electrons and photons exhibit low LET values up to 2 keV·µm$^{-1}$. The presence of titanium in the beam path attenuates their energy transfer, resulting in a noticeable reduction within the 1-2 keV·µm$^{-1}$. The insertion of titanium implants results in a significant increase in the LET spectra of ions, particularly within the 4-6.5 keV·µm$^{-1}$ range. This enhancement suggests an increased energy deposition from heavier ions, likely due to nuclear interactions with the titanium material. These interactions considerably modify the radiation field's composition and the energy transfer dynamics, indicating a more complex distribution of ionizing particles in the presence of high-Z materials like titanium.

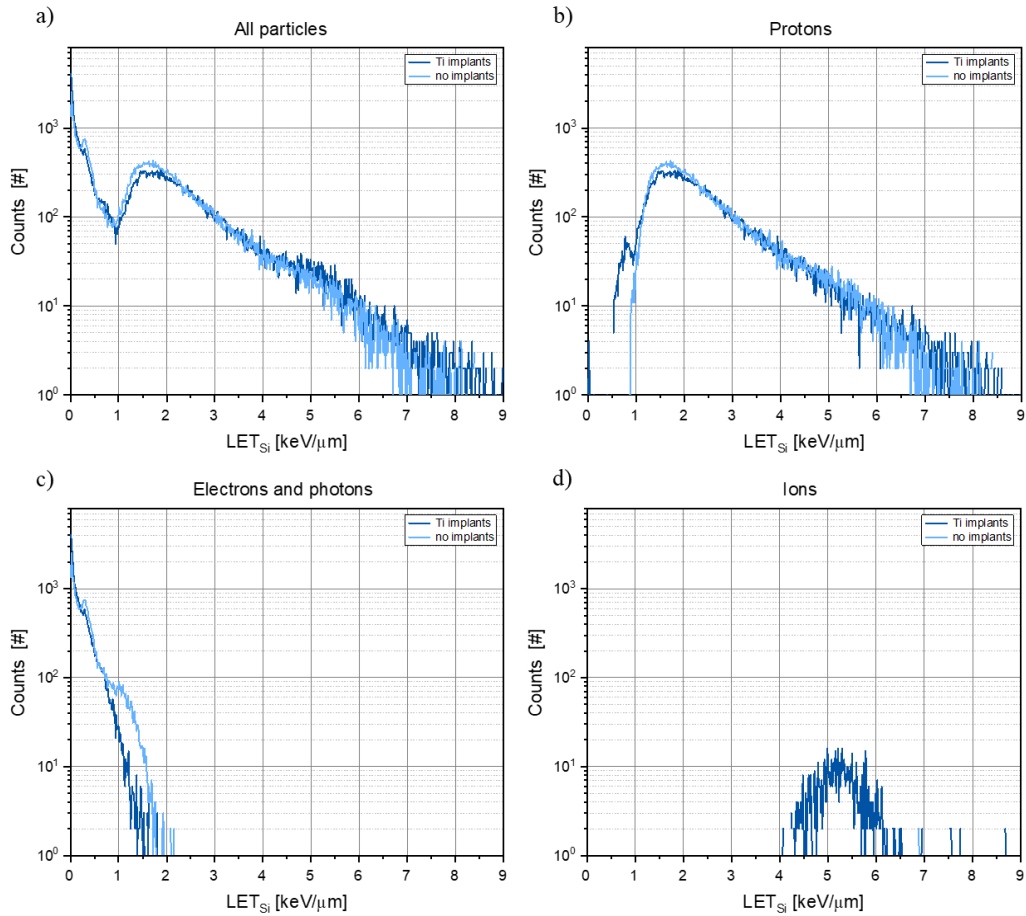

**Figure 4.** LET spectra of scattered radiation as measured by the Timepix3 detectors, comparing setups with and without titanium implants. The spectra are presented for: a) all particles, b) protons, c) electrons and photons and d) ions. Dark blue lines represent the setup with Ti implants, while light blue lines correspond to the setup without Ti implants in front of a 170 MeV clinical proton beam.

## 4. Conclusions

This work evaluates the impact of the Ti implants in proton therapy using high-resolution Minipix Timepix3 detectors in a clinical framework. Particle identification algorithms trained based on AI NN were used to decompose the radiation field into three groups of particles: (i) protons, (ii) electrons & photons and (iii) ions. Through detailed dosimetric analysis, including particle fluxes, dose rates (DR), and LET spectra, we observed distinct interactions between the proton beam and the Ti implants. The presence of titanium modifies the composition of secondary created radiation, leading to an increase in secondary ion production. Despite the notable increase in secondary ion dose rates with the implants, primary protons remain the dominant contributors to the overall dose deposition. Detailed LET spectra analysis shows that overall LET spectra with/without implants follows the same shape. At low-LET interval, below 1.5 keV·µm$^{-1}$, a shift was detected, indicating complex interactions between the proton beam and Ti. The results underscore the need for precise treatment planning in proton therapy for patients with metallic implants to ensure optimal dose delivery and minimize risks to surrounding healthy tissue. This study enhances our understanding of the interactions between proton beams and metallic implants, contributing to the ongoing development of safe and effective proton therapy protocols. Further research incorporating detailed treatment-based calculations and individualized patient assessments will be needed for advancing the accuracy and safety of proton therapy.

## Acknowledgements

The financial support for this experiment was done from project number 04-2-1132-2017/2022, founded from the agreement between JINR and Romania. Work at ADVACAM was performed in the frame of Contract No. 4000130480/20/NL/GLC/hh from the European Space Agency for the DPE project.